\documentclass[aps,prd,onecolumn,nofootinbib,superscriptaddress,12pt]{revtex4-2}

\usepackage{amsmath,amssymb,amsfonts,bm,mathtools}
\usepackage{graphicx}
\usepackage{hyperref}
\usepackage{xcolor}

\usepackage{booktabs}
\usepackage{microtype}
\usepackage{enumitem}
\allowdisplaybreaks

\newcommand{\mpl}{M_{\rm Pl}}
\newcommand{\Pz}{\mathcal P_\zeta}
\newcommand{\bk}{\bm k}
\newcommand{\bx}{\bm x}
\newcommand{\bq}{\bm q}

\usepackage{tikz}
\usetikzlibrary{arrows.meta,positioning,calc}
\begin{document}
\begin{flushright}
KOBE-COSMO-26-11
\end{flushright}

\title{Quasicrystalline Inflation:\\ From the Cosmological Dipole to Primordial Diffraction}

\author{Jiro Soda}
\affiliation{Department of Physics, Kobe University, Kobe 657-8501, Japan}
\date{November 4, 2026}

\begin{abstract}
We propose quasicrystalline inflation.   Six quasiperiodic phases decompose into three phonons and three phasons.  It turns out that a coherent pure-phason displacement can generate a local power dipole while the linear traceless quadrupole vanishes.  From spatially varying quadratic effective field theory, we derive off-diagonal correlations whose momentum transfer is restricted to integer combinations of the six icosahedral wavevectors. These correlated off-diagonal signals provide a characteristic observational signature of primordial quasiperiodic order, primordial diffraction.
\end{abstract}

\maketitle

\section{Introduction}
\label{sec:intro}

The statistical homogeneity and isotropy of primordial fluctuations are the cornerstones of the empirical success of inflation. Their possible small violations are nevertheless well-motivated observables because they probe spatial symmetries of the inflationary state. A standard parametrization of rotational-symmetry breaking is the preferred-direction spectrum~\cite{Ackerman2007},
\begin{equation}
P_\zeta(\bm k)=P_0(k)\left[1+g_*(\hat{\bm k}\!\cdot\!\hat{\bm n})^2+\cdots\right],
\label{eq:ACW}
\end{equation}
which preserves statistical homogeneity but violates statistical isotropy. A concrete and dynamically controlled realization is anisotropic inflation driven by a vector field with an inflaton-dependent kinetic function \cite{Watanabe2009,Kanno2010,Watanabe2010,Gumrukcuoglu2010,Dulaney2010}. Its perturbation theory, characteristic scalar--tensor cross-correlation, and CMB temperature/polarization signatures have been developed in detail \cite{Watanabe2010,Watanabe2011}; the subject and its generic predictions are reviewed in Ref.~\cite{Soda2012,Maleknejad:2012fw}. In this class, a homogeneous preferred direction primarily produces a quadrupolar statistical anisotropy in momentum space.

A conceptually different anomaly is the large-angle hemispherical power asymmetry (HPA) reported in WMAP and Planck data, which is commonly represented as a dipolar modulation in \emph{position space},
\begin{equation}
P_\zeta(k,\bm x)=P_0(k)\left[1+2A(k)\frac{\hat{\bm p}\cdot\bm x}{x_*}+\cdots\right].
\label{eq:HPAintro}
\end{equation}
Early WMAP analyses found that a dipole-modulated sky can fit the large-angle data better than a statistically isotropic sky \cite{Eriksen2007}. Planck's final isotropy analysis confirmed the persistence of several large-angle temperature anomalies while emphasizing their modest global significance and the limitations of polarization tests \cite{Planck2020}. Planck Pablic Release 4 temperature and polarization analyses continue to investigate the HPA, with the polarization evidence remaining weaker and sensitive to masks and estimator choices \cite{GimenoAmo2023,Sanyal2026,Lynch2026}. The observational situation is therefore intriguing but not decisive. 

Inflationary explanations of Eq.~\eqref{eq:HPAintro} often invoke a mode whose wavelength is much larger than our observable region. It is shown that a single adiabatic long mode is strongly constrained by homogeneity, whereas an additional field such as a curvaton can modulate short-scale power more efficiently \cite{Erickcek2008a,Erickcek2008b}. Isocurvature and multifield realizations can make the modulation scale dependent \cite{Erickcek2009};  non-Gaussian response \cite{Kanno:2013ohv} or non-vacuum initial states provide further possibilities \cite{Firouzjahi2014}. These works establish the long-mode modulation paradigm, but the modulating field is usually characterized by its long wavelength rather than by a distinctive spatial order.

Another line of development replaces a single preferred vector by an inflationary medium. Solid inflation uses three scalar condensates to break spatial diffeomorphisms while maintaining a homogeneous and isotropic background \cite{Endlich2013}. Such media can retain anisotropic deformations for an anomalously long time and can acquire a stochastic statistical anisotropy from superhorizon modes \cite{Bartolo2013,Bartolo2014}. Icosahedral inflation goes further:  a discrete icosahedral symmetry can keep the homogeneous background and the scalar two-point function isotropic while allowing anisotropy in higher-order observables~\cite{KangNicolis2016}.
It can be regarded as a realization of the general prediction in \cite{Yokoyama:2008xw}. These results show that nontrivial spatial order during inflation need not reduce to a single preferred vector.

There is also a useful bridge between form-field anisotropic inflation and inflationary media.  Two-form hair coupled to the inflaton provides a second controlled realization of anisotropic inflation~\cite{OhashiSodaTsujikawa2013,OhashiSodaTsujikawa2013Obs}.   It is emphasized that the apparently homogeneous two-form configuration is an inhomogeneous linear profile closely related to the construction underlying solid inflation \cite{ItoSoda2015}.  The relation has since been sharpened at the effective field theory (EFT) level~\cite{Gong:2019hwj}~\cite{AokiJimenezFigueruelo2023}.  This connection is especially relevant here because quasicrystalline inflation should be viewed not as unrelated to the form-field/solid family, but as an extension of the same symmetry logic from three translational Goldstone directions to a six-phase quasiperiodic structure with additional phason directions.

The missing ingredient that motivates the present work is \emph{quasiperiodic translational order}. Quasicrystals were experimentally identified through sharp diffraction peaks with noncrystallographic icosahedral symmetry and no Bravais-lattice translational periodicity \cite{Shechtman1984}; the theoretical notion of quasiperiodic order was formulated in \cite{LevineSteinhardt1984}. Their elasticity contains, in addition to ordinary phonons, internal phase modes called phasons \cite{LevineLubensky1985,Lubensky1985}. Modern EFT treatments organize these modes directly from symmetry principles \cite{BaggioliLandry2020}. Cosmological metrics with ``space-time quasicrystal'' structure have also been explored in a rather different nonholonomic/off-diagonal-gravity framework \cite{Vacaru2018}. To our knowledge, however, phasons of a primordial quasiperiodic medium have not been used as a mechanism for CMB hemispherical power modulation, nor has a rank-six quasiperiodic set of reciprocal wavevectors been developed as an inflationary two-point observable.  We distinguish this claim from earlier ``space-time quasicrystal'' cosmologies~\cite{Vacaru2018,AschheimVacaru2018}.

This paper combines these threads. Quasicrystals possess quasiperiodic translational order and phason modes in addition to phonons. We apply this structure to an inflationary phase medium and derive its primordial covariance. These phasons are absent in ordinary icosahedral solid inflation.

The paper is organized as follows. In section~\ref{sec:geometry}, we  introduce the quasicrystal order,  the icosahedral reciprocal star and its phonon--phason decomposition, with explicit contact to quasicrystal elasticity. 
We also introduce primordial diffraction. In section~\ref{sec:eft}, we give a background-compatible inflationary EFT and a simple positive-definite spectator action. In section~\ref{sec:power}, we show that the local cosmological dipole can be realized by specifying a phason configuration. In that case, the linear quadrapole anisotropy vanishes, and we show that the nonlinear quadrapole anisotropy is controlable. In section~\ref{sec:bragg}, we derive the primordial off-diagonal covariance, the primordial diffraction.
The final section is devoted to the conclusion.
In an appendix, we explain technical derivations.

\section{Icosahedral quasicrystalline order}
\label{sec:geometry}

\subsection{What is quasicrystalline order?}
\label{sec:qc_review}

Because quasicrystals are less familiar in cosmology than ordinary solids, we review the facts used.  A quasicrystal has long-range order and sharp diffraction, but its translational order is quasiperiodic rather than periodic.  The experimental discovery of an Al--Mn phase with sharp icosahedral diffraction peaks incompatible with a three-dimensional Bravais lattice established this possibility~\cite{Shechtman1984}. Subsequently, the quasicrystals as ordered structures with quasiperiodic translational order are formulated~\cite{LevineSteinhardt1984}.  Standard geometric and diffraction treatments may be found in Refs.~\cite{Senechal1995,BaakeGrimm2013,BaakeGrimm2012}.

For an ordinary three-dimensional crystal every Bragg vector can be written using three primitive reciprocal vectors,
\begin{equation}
 {\bf G}=\sum_{a=1}^{3} n_a {\bf b}_a,\qquad n_a\in{\mathbb Z}.
 \label{eq:crystal_module}
\end{equation}
 In a quasicrystal, more than three independent vectors may be required even in three physical dimensions. In our icosahedral construction, six vectors are needed.
Let
\begin{equation}
\tau=\frac{1+\sqrt5}{2},\qquad {\cal N}=\sqrt{1+\tau^2}=\sqrt{\tau+2},
\end{equation}
and choose six unit vectors along opposite pairs of icosahedral vertices,
\begin{align}
\bq_1&=\frac{1}{{\cal N}}(0,1,\tau), &
\bq_2&=\frac{1}{{\cal N}}(0,1,-\tau),&
\bq_3&=\frac{1}{{\cal N}}(1,\tau,0), \\
\bq_4&=\frac{1}{{\cal N}}(1,-\tau,0),&
\bq_5&=\frac{1}{{\cal N}}(\tau,0,1), &
\bq_6&=\frac{1}{{\cal N}}(\tau,0,-1).
\end{align}
The associated physical reciprocal vector is
\begin{equation}
 {\bf G}_{\bm n}=K\sum_{A=1}^{6}n_A\bq_A \ , \qquad
 \bm n\in\mathbb Z^6  \ .
\label{eq:fullmodule}
\end{equation}
Because the relations among the $\bq_A$ involve the irrational number $\tau$, this is not a simple rank-three reciprocal lattice. The point is that it has no nonzero integer kernel.  In fact, writing $\sum_A q^i_A n_A =0$ componentwise gives
\begin{align}
 n_3+n_4+\tau(n_5+n_6) =0,\qquad
 n_1+n_2+\tau(n_3-n_4) =0, \qquad
 \tau(n_1-n_2)+(n_5-n_6) =0.
\end{align}
Irrationality of $\tau$ forces rational and irrational coefficients to vanish separately, so $n_A=0$ for all $A$.
We refer to this integer-generated set as the rank-six reciprocal module. 
The excess rank is the origin of the additional internal phase degrees of freedom.   Nevertheless the structure is ordered: ideal model sets can exhibit pure-point diffraction with sharp Bragg peaks \cite{BaakeGrimm2012,BaakeGrimm2013}.  This coexistence of sharp diffraction and nonperiodicity is the property we exploit cosmologically.

\subsection{Phonons, phasons, and elasticity}
\label{sec:phonon_phason_review}

We regard the six $\Theta_A$ as compact torus phase fields,
\begin{equation}
 \Theta_A \sim \Theta_A+2\pi .
\end{equation}
In the ideal phase medium, the leading low-energy theory is invariant
under continuous shifts $\Theta_A\to\Theta_A+c_A$.  The background
$\bar\Theta_A=K\bm q_A\cdot\bm x+\phi_A$ spontaneously selects a point
in this phase space.  Of the six corresponding phase directions,
three are generated by physical translations,
\begin{equation}
 \delta\phi_A=K\,\bm q_A\cdot\bm u ,
\end{equation}
while the three independent directions satisfying
\begin{equation}
 \sum_{A=1}^{6}\bm q_A\,\delta\phi_A=0
\end{equation}
are phasons.  Thus the ideal rank-six phase medium contains three
phonon and three phason Goldstone directions.
The continuous phase shifts refer to the ideal spectator sector. The phase-dependent curvature portal introduced below provides an explicit weak breaking, so the phasons should more precisely be regarded as approximate Goldstone modes in the full EFT.
For later use we define the projectors onto these two subspaces by
\begin{equation}
 (\Pi_{\parallel})_{AB}\equiv
 \frac12\,\bm q_A\cdot\bm q_B,
 \qquad
 (\Pi_{\perp})_{AB}\equiv
 \delta_{AB}-(\Pi_{\parallel})_{AB}.
\end{equation}
One can verify
$\Pi_{\parallel}^2=\Pi_{\parallel}$,
$\Pi_{\perp}^2=\Pi_{\perp}$, and
$\Pi_{\parallel}\Pi_{\perp}=0$.
Taking a convenient basis in a phason space
\begin{align}
P^{(1)}=(\tau,\tau,-1,1,0,0)^T \ ,\qquad
P^{(2)}=(1,\tau,-\tau,0,1,0)^T \ , \qquad 
P^{(3)}=(\tau,1,-\tau,0,0,1)^T \ ,
\end{align}
we can separate three translation/phonon coordinates from three perpendicular-space phasons as
\begin{equation}
 \delta\phi_A=Kq_{Ai}u^i+P_A^{(a)}w_a \ .
\label{eq:phononphason}
\end{equation}
The field $u^i$ is the ordinary phonon displacement: a uniform $u^i$ is equivalent to translating the pattern in physical space.  The field $w^a$ is the phason displacement: it changes the relative phases of the six quasiperiodic phases and corresponds geometrically to motion in $E_\perp$.  Thus a phason is not an additional acoustic polarization of an ordinary crystal.  It exists because the rank of quasiperiodic order exceeds the physical dimension.
This is the standard kinematical content of icosahedral quasicrystal elasticity \cite{LevineLubensky1985,Lubensky1985,deBoissieu2012,BaggioliLandry2020}; modern dual formulations likewise treat phonon and phason Goldstone sectors separately \cite{Surowka2021}.

The long-wavelength elasticity of pentagonal and icosahedral quasicrystals was formulated soon after their discovery \cite{LevineLubensky1985,Lubensky1985}.  Schematically, the static elastic free energy contains
\begin{equation}
 F_{\rm el}
 =\frac12\int d^3x\,
 \left[
 C_{ijkl}u_{ij}u_{kl}
 +K_{abij}(\partial_iw^a)(\partial_jw^b)
 +2R_{ijka}u_{ij}\partial_k w^a
 \right],
 \label{eq:qc_elasticity_review}
\end{equation}
where $u_{ij}=(\partial_i u_j+\partial_j u_i)/2$ and $C_{ijkl}, K_{abij}, R_{ijka}$ are coupling functions.  The last term describes phonon--phason coupling.  Microscopic and experimental studies show that such coupling can be nonzero, while the phason elastic sector controls characteristic diffuse scattering around Bragg peaks \cite{ZhuHenley1999,deBoissieu2012}.  This is useful conceptually for the present model: the cosmological phason is an internal deformation of quasiperiodic order, not an ad hoc spectator label.

There is, however, an important dynamical distinction between laboratory quasicrystals and the inflationary EFT used here.  In finite-temperature hydrodynamics the asymptotically long-wavelength phason is generically diffusive, a classic prediction of quasicrystal hydrodynamics that is supported by scattering measurements \cite{Lubensky1985,deBoissieu2012}.  Modern Schwinger--Keldysh EFT makes clear how this dissipative behavior follows from the symmetry structure at finite temperature \cite{BaggioliLandry2020}.  We do \emph{not} import that dissipative constitutive relation into inflation.  Our primordial medium is treated as a weakly coupled relativistic spectator during a quasi-de Sitter epoch, and its kinetic terms determine whether the phason is propagating, overdamped, or weakly pinned.  What is borrowed from quasicrystal physics is the symmetry and kinematics of the phonon--phason decomposition.

\subsection{Diffraction and the cosmological analogy}
\label{sec:qc_diffraction_review}

  In an icosahedral quasicrystal the standard superspace description is six-dimensional, with $\mathbb R^6=E_\parallel^3\oplus E_\perp^3$; this description is used both in structural refinements and in the analysis of phason diffuse scattering \cite{vanSmaalen1991,Yamada2016}.  
This is the precise sense in which we use the term \emph{primordial diffraction}.  We do not identify the inflationary state with an atomic lattice.  We assume that a spectator order parameter is quasiperiodic in three physical dimensions and is periodic on a six-dimensional phase torus.    The twelve vectors $\pm K{\bf q}_A$ form the leading first star, while higher integer combinations give weaker diffraction components.

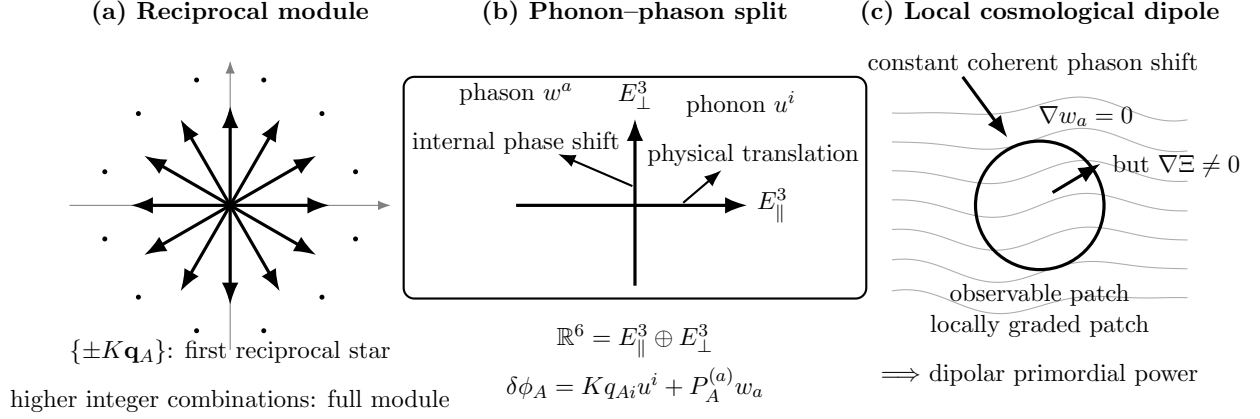
\begin{figure*}[t]
\centering
\begin{tikzpicture}[font=\small,>=Latex,scale=0.85,transform shape]
\begin{scope}[xshift=0cm]
\node[font=\bfseries] at (0,3.0) {(a) Reciprocal module};
\draw[->,gray] (-2.5,0) -- (2.5,0);
\draw[->,gray] (0,-2.25) -- (0,2.25);
\foreach \ang in {0,30,...,330}{\draw[very thick,->] (0,0) -- ({1.55*cos(\ang)},{1.55*sin(\ang)});}
\foreach \ang in {15,45,...,345}{\fill ({2.02*cos(\ang)},{2.02*sin(\ang)}) circle (1.1pt);}
\node[align=center] at (0,-2.65) {$\{\pm K{\bf q}_A\}$: first reciprocal star\\higher integer combinations: full module};
\end{scope}
\begin{scope}[xshift=6.3cm]
\node[font=\bfseries] at (0,3.0) {(b) Phonon--phason split};
\draw[rounded corners,thick] (-3.6,-1.45) rectangle (3.6,2.0);
\draw[->,very thick] (-1.85,0) -- (1.75,0) node[right] {$E_\parallel^3$};
\draw[->,very thick] (0,-1.25) -- (0,1.35) node[above] {$E_\perp^3$};
\draw[->,thick] (0.75,0.04) -- (1.35,0.55);
\node[align=center] at (1.65,1.2) {phonon $u^i$\\\quad physical translation};
\draw[->,thick] (-0.04,0.3) -- (-1.2,0.8);
\node[align=center] at (-1.85,1.35) {phason $w^a$\\internal phase shift};
\node[align=center] at (0,-2.45) {$\mathbb R^6=E_\parallel^3\oplus E_\perp^3$\\$\delta\phi_A=Kq_{Ai}u^i
+P_{A}^{(a)} w_a$};
\end{scope}
\begin{scope}[xshift=12.6cm]
\node[font=\bfseries] at (0,3.0) {(c) Local cosmological dipole};
\foreach \y in {-1.5,-1.0,...,1.5}{
  \draw[gray!65] plot[smooth,samples=50,domain=-2.3:2.3] (\x,{\y+0.13*sin(95*\x+35*\y)+0.07*sin(155*\x-20*\y)});
}
\draw[very thick] (0,0) circle (1.0);
\node at (0,-1.4) {observable patch};
\draw[->,very thick] (-1.2,2.0) -- (-0.5,1.05);
\node[align=left] at (-0.1,2.2) {constant coherent phason shift};
\draw[->,very thick] (0.2,0.2) -- (0.95,0.65);
\node[align=left] at (1.55,1.0) {$\nabla w_a=0$\\\quad\qquad but $\nabla\Xi\neq0$};
\node[align=center] at (0,-2.25) {locally graded patch\\$\Longrightarrow$ dipolar primordial power};
\end{scope}
\end{tikzpicture}
\caption{\label{fig:qc_concept}
Quasicrystalline order, phasons, and the cosmological dipole. (a) The twelve vectors $\{\pm K{\bf q}_A\}$ form the first reciprocal star, the lowest-index icosahedral point-group orbit; integer combinations generate the full rank-six Fourier module. The outer points are schematic and indicate higher-index module elements rather than a literal projection of the full three-dimensional module. (b) The six-dimensional phase space decomposes into three physical translation (phonon) directions $E_\parallel$ and three perpendicular (phason) directions $E_\perp$. (c) A coherent spatially constant phason displacement moves the state through  a different quasiperiodic phase configuration and can select an observable patch with a nonzero physical-space gradient of the order parameter. Hence $\nabla w_a=0$ can coexist with $\nabla\Xi\neq0$, producing a local dipolar modulation.}
\end{figure*}

We introduce the first-star scalor profile
\begin{equation}
\Xi({\bf x})=\sum_A\cos\Theta_A({\bf x}).
\label{eq:Xi}
\end{equation}
In the minimal EFT the higher Fourier amplitudes are not independent dynamical density-wave fields; the full hierarchy is generated by an analytic periodic observable of the six quasiperiodic phases.  Thus this is a quasiperiodic phase EFT with a calculable nonlinear diffraction hierarchy, not a microscopic derivation of independently stabilized amplitudes on every module vector.  The analytic periodic response used in Sec.~\ref{sec:eft},
\begin{equation}
 {\cal E}(\bm\Theta)\equiv e^{\beta\Xi(\bm\Theta)},
\label{eq:Egenerator}
\end{equation}
has the exact Fourier expansion
\begin{equation}
 {\cal E}(\bm\Theta)=
 \sum_{\bm n\in\mathbb Z^6}
 \left[\prod_{A=1}^{6}I_{n_A}(\beta)\right]
 e^{i\bm n\cdot\bm\Theta},
\label{eq:Besselexpansion}
\end{equation}
where $I_n$ is the modified Bessel function of the first kind.  Hence, the model has support on every \(\mathbf G_{\mathbf n}\) defined in Eq. (7). For $\beta\ll1$, $I_1/I_0\simeq\beta/2$ and higher-index coefficients are parametrically suppressed, so the twelve-vector calculation is the controlled \emph{first-star limit} of a genuine rank-six quasiperiodic response rather than its definition.

\section{Quasicrystalline inflationary effective theory}
\label{sec:eft}

A fully microscopic relativistic quasicrystal is not required for the two-point phenomenology, but the EFT should distinguish the parallel and perpendicular phase sectors.  We use
\begin{align}
S=&\int d^4x\sqrt{-g}\left[
\frac{\mpl^2}{2}R-\mpl^2(3H^2+\dot H)+\mpl^2\dot H g^{00}
+\frac12M_2^4(\bm\Theta)(\delta g^{00})^2+\ldots\right]
+S_{\rm QC},
\label{eq:EFT}
\end{align}
where $\delta g^{00}=g^{00}+1$.  Because this operator vanishes on the unperturbed FRW background, its quasiperiodic coefficient can modulate scalar perturbations without directly sourcing a comparably large background density modulation.

We retain the analytic phase response
\begin{equation}
 M_2^4(\bm\Theta)=\bar M_2^4
 \exp\{\beta[\Xi(\bm\Theta)-\Xi_0]\},
\label{eq:M2}
\end{equation}
where
\begin{equation}
 \Xi_0\equiv \Xi({\bf x}=0)=\sum_{A=1}^{6}\cos\phi_A
\label{eq:Xi0definition}
\end{equation}
is the value at the center of the observable patch.  
More generally, because the $\Theta_A$ are compact phases, any
phase-dependent EFT coefficient is a periodic function and can be
expanded as
\begin{equation}
 F(\Theta)=
 \sum_{\bm n\in\mathbb Z^6}
 F_{\bm n}\,e^{i\bm n\cdot\bm\Theta}.
\end{equation}
Evaluated on
$\bar\Theta_A=K\bm q_A\cdot\bm x+\phi_A$, this becomes
\begin{equation}
 F(\bar\Theta)=
 \sum_{\bm n}
 F_{\bm n}\,
 e^{i\bm n\cdot\bm\phi}
 e^{i\bm G_{\bm n}\cdot\bm x},
 \qquad
 \bm G_{\bm n}=K\sum_A n_A\bm q_A .
\end{equation}
Thus the quasiperiodic spatial dependence of the EFT coefficients,
and hence the momentum-transfer selection rule derived below, follows
from evaluating periodic functions of the compact phases on the
quasiperiodic phase background.  Eq.~(\ref{eq:M2}) is a simple analytic
benchmark choice within this general structure.
This analytic response is useful because Eq.~\eqref{eq:Besselexpansion} makes its full rank-six content explicit.  Equation~\eqref{eq:M2} is the perturbative portal between the quasiperiodic spectator and the adiabatic mode.  It is not derived from $S_{\rm QC}$ itself.  In unitary gauge, phase-dependent operators such as $F_1(\bm\Theta)\delta g^{00}$, $F_2(\bm\Theta)\delta K$, and $F_3(\bm\Theta)(\delta g^{00})^2$ are allowed subject to the same phase periodicity.  We use the last operator because it vanishes on the FRW background and changes the scalar kinetic coefficient without introducing a leading background force.  Other portals modify response kernels but preserve the reciprocal-module selection rule whenever their coefficients are periodic analytic functions of the six phases.

We denote the inflaton by $\varphi$; it controls the slowly varying
normalizations $f_{\parallel}(\varphi)$ and
$f_{\perp}(\varphi)$.
For the spectator sector, we choose a projector-resolved positive definite kinetic action  in the more general quasicrystal EFT parameter space,
\begin{align}
S_{\rm QC}=-\frac12\int d^4x\sqrt{-g}\,
\Big[&f_\parallel^2(\varphi)(\Pi_\parallel)_{AB}
+f_\perp^2(\varphi)(\Pi_\perp)_{AB}\Big]
 g^{\mu\nu}\partial_\mu\Theta_A\partial_\nu\Theta_B
\ .
\label{eq:SQprojected}
\end{align}
The background gradient $\partial_i\Theta_A=Kq_{Ai}$ lies entirely in the parallel sector, so only $f_\parallel$ is required to track the background.  Taking
\begin{equation}
 \frac{d\ln f_\parallel}{dN}=1+\delta_f,
 \qquad |\delta_f|\ll1,
\label{eq:ftracking}
\end{equation}
gives $\rho_{\rm QC}\simeq f_\parallel^2 K^2 /a^2 \propto e^{2\delta_fN}$ and can keep the spectator fraction approximately constant during inflation, in direct analogy with the kinetic-function mechanism used for inflationary vector and two-form hair \cite{Watanabe2009,OhashiSodaTsujikawa2013,Soda2012}.  The phason normalization $f_\perp$ is independent because a homogeneous phason displacement carries no background gradient energy. 

Equation~\eqref{eq:SQprojected} is the simplest positive-definite member of a wider $F(B^{AB})$ theory, $B^{AB}=g^{\mu\nu}\partial_\mu\Theta_A\partial_\nu\Theta_B$.  Generic quasicrystal elasticity allows distinct phonon and phason moduli and phonon--phason mixing \cite{LevineLubensky1985,ZhuHenley1999,Yamada2016}.  We set the mixing to zero at the benchmark point rather than claiming it is forbidden.  Small symmetry-allowed mixing can be restored perturbatively without changing the rank-six selection rule.

The curvature perturbation has usual local quadratic form, generalized to a slowly varying spatial coefficient 
\begin{equation}
S_\zeta^{(2)}=\int dt\,d^3x\,a^3\frac{\epsilon\mpl^2}{c_s^2({\bf x})}
\left[\dot\zeta^2-c_s^2({\bf x})\frac{(\nabla\zeta)^2}{a^2}\right],
\label{eq:zetaaction}
\end{equation}
with
\begin{equation}
 c_s^{-2}({\bf x})=1+u\,e^{\beta\Delta\Xi({\bf x})},\qquad
 u=\frac{2\bar M_2^4}{\epsilon\mpl^2H^2},
\label{eq:cs}
\end{equation}
and $\Delta\Xi=\Xi-\Xi_0$.  The local reference sound speed is therefore
\begin{equation}
 c_{s,{\rm loc}}^{-2}\equiv c_s^{-2}({\bf 0})=1+u.
\label{eq:cslocal}
\end{equation}
This local value, used in the horizon-crossing calculation of Sec.~IV, is distinct from the Fourier zero mode $\bar c_s^{-2}$ of the complete quasiperiodic pattern introduced in Sec.~V.  The local/separate-universe expansion requires $K\ll k$ for the short curvature mode.  The full module itself is not truncated by this approximation; only its local gradient expansion is.  Section IV uses this local limit to extract the dipole and its angular structure. Section V instead treats the spatial Fourier components of the quadratic EFT directly.

\subsection{Background solution and the action for curvature perturbations}
\label{sec:background_derivation}

It is useful to separate the existence of the spatial background from the question of its dynamical formation.  For a general covariant phase-medium action
\begin{equation}
 S_F=\int d^4x\sqrt{-g}\,F(B^{AB},\varphi),
 \qquad
 B^{AB}=g^{\mu\nu}\partial_\mu\Theta_A\partial_\nu\Theta_B ,
\label{eq:FBbackground}
\end{equation}
the phase equations are
\begin{equation}
 \nabla_\mu\!\left(2F_{AB}\nabla^\mu\Theta_B\right)=0,
 \qquad F_{AB}\equiv\frac{\partial F}{\partial B^{AB}} .
\label{eq:phaseeom}
\end{equation}
On an FRW geometry, the ansatz
\begin{equation}
 \bar\Theta_A=Kq_{Ai}x^i+\phi_A
\label{eq:backgroundsolution}
\end{equation}
has $B^{AB}=K^2a^{-2}\,\bq_A\!\cdot\!\bq_B$, which is spatially constant.  Hence $F_{AB}$ depends only on time and Eq.~\eqref{eq:phaseeom} is satisfied identically.  If $F$ respects the icosahedral symmetry, the spatial stress is isotropic because its rank-two part is proportional to
$\sum_Aq_A^iq_A^j=2\delta^{ij}$.  Thus a spatially nontrivial rank-six phase background is compatible with an FRW metric.  The tracking choice in Eq.~\eqref{eq:ftracking} is needed to prevent dilution of its energy density, not for the existence of the solution itself.  

For completeness we also derive Eq.~\eqref{eq:zetaaction}.  Restoring the Goldstone mode $\pi$ by the Stückelberg replacement $t\to t+\pi$ gives on FRW
\begin{equation}
 \delta g^{00}=-2\dot\pi-\dot\pi^2+\frac{(\nabla\pi)^2}{a^2}+\cdots,
 \qquad
 (\delta g^{00})^2=4\dot\pi^2+O(\pi^3).
\label{eq:stueckelberg}
\end{equation}
The quadratic decoupling-limit action following from Eq.~\eqref{eq:EFT} is therefore
\begin{align}
 S_\pi^{(2)}
 =\int dt\,d^3x\,a^3\Bigg[
 &\left(\epsilon\mpl^2H^2+2M_2^4(\bm\Theta)\right)\dot\pi^2
 -\epsilon\mpl^2H^2\frac{(\nabla\pi)^2}{a^2}
 \Bigg].
\label{eq:piaction}
\end{align}
Defining
\begin{equation}
 c_s^{-2}({\bf x})=
 1+\frac{2M_2^4(\bm\Theta({\bf x}))}{\epsilon\mpl^2H^2}
\label{eq:csderive}
\end{equation}
and using $\zeta=-H\pi$ at leading slow-roll order yields Eq.~\eqref{eq:zetaaction}.  Because $M_2^4$ varies in space, this is the leading local action in a gradient expansion.  Besides $K\ll k$, the more explicit adiabatic condition is
\begin{equation}
 \left|\frac{\partial_i c_s}{k\,c_s}\right|\ll1,
\label{eq:adiabaticcs}
\end{equation}
and omitted terms are suppressed by gradients of the quasiperiodic coefficients, parametrically by powers of $K/k$.

\section{Cosmological dipole}
\label{sec:power}

For $K\ll k$, a short mode sees an approximately constant local sound speed at horizon crossing,
\begin{equation}
 \Pz(k,{\bf x})=\frac{H_k^2}{8\pi^2\epsilon_k\mpl^2c_s({\bf x})},
\end{equation}
and hence
\begin{equation}
 \frac{\Pz(k,{\bf x})}{\Pz^{(0)}(k)}
 =\left[\frac{1+u e^{\beta\Delta\Xi({\bf x})}}{1+u}\right]^{1/2}.
\label{eq:exactresponse}
\end{equation}
At linear response, with the unmodulated spectrum in the denominator,
\begin{equation}
 \frac{\delta\Pz}{\Pz^{(0)}}=\gamma\Delta\Xi+O(\Delta\Xi^2),
 \qquad
 \gamma=\frac{\beta u}{2(1+u)}.
\label{eq:gamma}
\end{equation}
At first order one may equivalently write $\delta\Pz/\Pz$; the distinction matters only at the nonlinear order considered below.
Let $\kappa=Kx_*$ and ${\bf x}=x_*\hat{\bf n}$.  Expanding the first star locally,
\begin{align}
\Delta\Xi(\hat{\bf n})= -\kappa\sum_A\sin\phi_A(\bq_A\cdot\hat{\bf n})
-\frac{\kappa^2}{2}\sum_A\cos\phi_A(\bq_A\cdot\hat{\bf n})^2
+\frac{\kappa^3}{6}\sum_A\sin\phi_A(\bq_A\cdot\hat{\bf n})^3+O(\kappa^4).
\label{eq:gradexp}
\end{align}
When $|{\bf G}_{\bm n}|x_*\ll1$, each quasiperiodic harmonic is locally indistinguishable from a superhorizon long mode. The dipole alone therefore does not identify quasicrystalline order. The distinctive prediction is that all off-diagonal correlations share the same integer-generated reciprocal structure, phase relations, and amplitude hierarchy.

Writing
\begin{equation}
 \Pz(k,\hat{\bf n})=\Pz^{(0)}(k)
 [1+2{\bf A}(k)\cdot\hat{\bf n}+{\cal Q}_{ij}n^in^j+\cdots],
\end{equation}
gives
\begin{equation}
 A_i=-\frac{\gamma\kappa}{2}\sum_A\sin\phi_A q_{Ai},
\label{eq:dipolegeneral}
\end{equation}
\begin{equation}
 {\cal Q}_{ij}=-\frac{\gamma\kappa^2}{2}
 \left[C_{ij}-\frac13\delta_{ij}C_{kk}\right],\qquad
 C_{ij}=\sum_A\cos\phi_Aq_{Ai}q_{Aj}.
\label{eq:quadgeneral}
\end{equation}
Thus dipole dominance is generic in the long-pattern limit: away from accidental cancellations, $|{\cal Q}|/A=O(\kappa)$.

\subsection{An exactly pure-phason benchmark}
\label{sec:purebenchmark}

An exactly pure-phason benchmark is obtained by choosing a configuration 
\begin{equation}
 w_1=w_2=w_3=w_*,\qquad
 w_*\equiv\frac{\pi}{\tau^2},
\label{eq:wstar}
\end{equation}
and defining $\bm\phi=\sum_aw_*P^{(a)}$.  By construction
\begin{equation}
 \sum_{A=1}^{6}\bm q_A\,\phi_A=0,
 \qquad
 \Pi_\perp\bm\phi=\bm\phi ,
\end{equation}
so the displacement has no translation/phonon component and is
purely phasonic.
 Modulo $2\pi$, an equivalent phase representative is
\begin{equation}
 (\phi_1,\ldots,\phi_6)\equiv(-w_*,-w_*,+w_*,+w_*,+w_*,+w_*)\pmod{2\pi}.
\label{eq:purephases}
\end{equation}
All six cosines are then equal, and Eq.~\eqref{eq:quadgeneral} implies
\begin{equation}
 C_{ij}=2\cos w_*\,\delta_{ij} \ .
\end{equation}
Thus, we obtain
\begin{equation}
 {\cal Q}_{ij}=0,
 \label{eq:quadzero}
\end{equation}
at linear response.  With $s_A=(-1,-1,+1,+1,+1,+1)$,
\begin{equation}
 {\bf S}\equiv\sum_As_A\bq_A,
 \qquad |{\bf S}|^2=6+\frac{6\sqrt5}{5},
\end{equation}
and the dipole becomes
\begin{equation}
 A(k)=\chi\,\gamma(k)Kx_*,\qquad
 \chi\equiv\frac12\sin w_*
 \sqrt{6+\frac{6\sqrt5}{5}}\simeq1.373229 .
\label{eq:centralA}
\end{equation}
The exact cancellation is a special point, but dipole dominance is not.  Along the pure-phason line $w_1=w_2=w_3=w_*+\delta w$, direct expansion gives
\begin{equation}
 \frac{\sqrt{{\cal Q}_{ij}{\cal Q}_{ij}}}{A}
 =1.9464\,\kappa|\delta w|+O(\kappa\delta w^2,\kappa^2),
\label{eq:tuningmeasure}
\end{equation}
so a finite neighborhood of the benchmark remains strongly dipole dominated when $\kappa<1$.

A constant phason shift is sufficient: it selects a different  quasiperiodic phase configuration, and the selected phase generally has $\nabla\Xi\neq0$ across our finite patch.  The dipole therefore need not arise from $\nabla w\neq0$ itself, but from the local gradient of the quasiperiodic order selected by a constant perpendicular-space displacement.

\subsection{Nonlinear quadrupole}

The geometrical quadrupole in Eq.~\eqref{eq:quadzero} vanishes only at first order in the response.  Expanding Eq.~\eqref{eq:exactresponse},
\begin{equation}
 \frac{\Pz}{\Pz^{(0)}}=1+\gamma\Delta\Xi
 +\frac12(\eta+\gamma^2)(\Delta\Xi)^2+\cdots,
 \qquad
 \eta=\frac{\beta^2u}{2(1+u)^2}.
\end{equation}
For the pure-phason point, the anisotropic $O(\kappa^2)$ contribution comes from the square of the dipole.  If the position-space quadrupole is written as $g_2P_2(\hat{\bf p}\cdot\hat{\bf n})$, then
\begin{equation}
 g_2^{\rm nl}=\frac43\left(1+\frac{2}{u}\right)A^2 .
\label{eq:nonlinearquad}
\end{equation}
This coefficient describes a real-space $L=2$ modulation and is distinct from the homogeneous momentum-space ACW parameter $g_*$.   Planck finds no compelling evidence for primordial quadrupolar statistical anisotropy, while direction-dependent primordial reconstructions probe amplitudes at roughly the $10^{-2}$ scale \cite{Planck2018Inflation,Durakovic2018}.  
If we adopt
\begin{equation}
 \beta=0.30,\qquad u=5,\qquad \kappa=0.408,
\label{eq:balancedpoint}
\end{equation}
for which $\gamma=0.125$, $c_{s,{\rm loc}}=1/\sqrt6=0.408$, $A=0.0700$, then we get $g_2^{\rm nl}=9.15\times10^{-3}$.  Thus the nonlinear quadrupole can be reduced below one percent without losing the target dipole, although a dedicated $L=2$ likelihood remains necessary.

This benchmark applies to large scales. Small-scale CMB data constrain
$A<0.0045$ at roughly $10\,{\rm Mpc}$ (95\% C.L.)~\cite{FlenderHotchkiss2013}.
A possible extension is to let $\beta=\beta(\varphi)$ in
Eq.~\eqref{eq:M2}, keeping $u$ approximately constant. For modes leaving
the sound horizon during nearly constant-$\beta$ intervals,
\[
 A(k)\simeq\chi\kappa\,\frac{u\,\beta(t_k)}{2(1+u)},
 \qquad c_{s,{\rm loc}}k=a(t_k)H(t_k).
\]
With $u=5$ and $\kappa=0.408$, two plateaus with $\beta=0.30$ and
$0.018$ give $A\simeq0.0700$ and $0.0042$, respectively. The local
sound speed remains unchanged because $\Delta\Xi({\bf0})=0$.
The portal still vanishes on the FRW background and introduces no new
propagating field. A smooth transition can occur after the large-scale
modes freeze, while remaining adiabatic for smaller-scale subhorizon
modes. This is an illustrative EFT profile; predictions during the
transition require solving the time-dependent mode equation.

\section{Primordial diffraction}
\label{sec:bragg}

The selection rule follows directly from the spatially varying quadratic EFT.  Using Eq.~\eqref{eq:Besselexpansion}, the exact Fourier decomposition of the kinetic coefficient is
\begin{align}
 c_s^{-2}({\bf x})
 =\,1+u e^{-\beta\Xi_0}\sum_{\bm n\in\mathbb Z^6}
 e^{i\bm n\cdot\bm\phi}
 \left[\prod_{A=1}^{6}I_{n_A}(\beta)\right]
 e^{i{\bf G}_{\bm n}\cdot{\bf x}} 
 \equiv \,\bar c_s^{-2}+\delta C({\bf x}).
\label{eq:exactCdecomp}
\end{align}
The homogeneous coefficient and the nonzero module amplitudes are
\begin{equation}
 \bar c_s^{-2}=1+u e^{-\beta\Xi_0}I_0(\beta)^6,
 \qquad
 \delta C_{\bm n}=u e^{-\beta\Xi_0}
 e^{i\bm n\cdot\bm\phi}
 \prod_{A=1}^{6}I_{n_A}(\beta),\quad \bm n\neq0.
\label{eq:deltaCnexplicit}
\end{equation}
The Fourier zero mode $\bar c_s^{-2}$ is a phase-space average and is not identical to the local reference value $c_{s,{\rm loc}}^{-2}=1+u$.  At the preferred benchmark, $\Xi_0=6\cos(\pi/\tau^2)=2.17425$, so
\begin{equation}
 c_{s,{\rm loc}}=0.40825,\qquad
 \bar c_s=0.50135.
\label{eq:soundspeedbenchmark}
\end{equation}
Since $I_{-n}(\beta)=I_n(\beta)$ for integer $n$,
\begin{equation}
 \delta C_{-\bm n}=\delta C_{\bm n}^*,
\label{eq:realitydeltaC}
\end{equation}
which is the reality condition for $c_s^{-2}({\bf x})$.  Hence
\begin{equation}
 \delta C({\bf x})=\sum_{\bm n\neq0}\delta C_{\bm n}
 e^{i{\bf G}_{\bm n}\cdot{\bf x}} .
\label{eq:deltaCmodule}
\end{equation}
At first order in $\delta C$, Eq.~\eqref{eq:zetaaction} contains
\begin{equation}
 S_{\rm int}
 =\epsilon\mpl^2\int d\eta\,d^3x\,a^2(\eta)\,
 \delta C({\bf x})\,\zeta'^2 .
\label{eq:Sinint}
\end{equation}
The in-in correction is
\begin{align}
 \delta\langle\zeta_{\bf k}\zeta_{{\bf k}'}\rangle
 &=-i\int_{-\infty}^{0}d\eta\,
 \left\langle\left[
 \zeta_{\bf k}(0)\zeta_{{\bf k}'}(0),H_{\rm int}(\eta) 
 \right]\right\rangle 
 =2\ {\rm Im} \int_{-\infty}^{0}d\eta\,
 \left\langle
 \zeta_{\bf k}(0)\zeta_{{\bf k}'}(0) H_{\rm int}(\eta) 
 \right\rangle\\
 &=(2\pi)^3\sum_{\bm n\neq0}
 \delta^3({\bf k}+{\bf k}'-{\bf G}_{\bm n})\,
 \delta C_{\bm n}\,
 {\cal F}(k,k';{\bf G}_{\bm n}) .
\label{eq:ininmodule}
\end{align}
We use
\begin{equation}
 \zeta({\bf x})=\int\frac{d^3k}{(2\pi)^3}\,
 e^{i{\bf k}\cdot{\bf x}}\zeta_{\bf k},
\label{eq:Fourierconv}
\end{equation}
so that no Fourier factors are absorbed into $\delta C_{\bm n}$.  To first order in the kinetic interaction,
\begin{equation}
 {\cal F}(k,k')
 =-4\epsilon\mpl^2\,{\rm Im}\!
 \left[
 u_k(0)u_{k'}(0)
 \int_{-\infty(1-i0^+)}^{0}d\eta\,a^2
 u_k^{\prime *}(\eta)u_{k'}^{\prime *}(\eta)
 \right].
\label{eq:Fkernel}
\end{equation}
For constant $H$, $\epsilon$, and $\bar c_s$, the Bunch--Davies mode
\begin{equation}
 u_k(\eta)=\frac{H}{2\sqrt{\epsilon \bar c_s}\,\mpl k^{3/2}}
 (1+i \bar c_s k\eta)e^{-i \bar c_s k\eta}
\end{equation}
gives
\begin{equation}
 {\cal F}_{\rm dS}(k,k')
 =\frac{H^2\bar c_s}
 {4\epsilon\mpl^2\,k k'(k+k')}.
\label{eq:FdeSitter}
\end{equation}
Here
\begin{equation}
 P_0(k)\equiv |u_k(0)|^2
 =\frac{2\pi^2}{k^3}\,{\cal P}^{(0)}_\zeta(k).
\label{eq:P0definition}
\end{equation}
For $k'=k$, ${\cal F}_{\rm dS}/P_0=\bar c_s^{\,2}/2$, reproducing the local response to a perturbation around the homogeneous coefficient $\bar c_s^{-2}$.  Appendix~\ref{app:inin} derives the Legendre transform, the interaction-Hamiltonian sign, and the in-in response kernel.

Combining Eqs.~\eqref{eq:deltaCnexplicit} and \eqref{eq:ininmodule} gives the leading-order primordial-diffraction formula
\begin{equation}
 \begin{aligned}
 \delta\langle\zeta_{\bf k}\zeta_{{\bf k}'}\rangle
 =(2\pi)^3 u e^{-\beta\Xi_0}
 \sum_{\bm n\neq0}e^{i\bm n\cdot\bm\phi}
 \left[\prod_{A=1}^{6}I_{n_A}(\beta)\right]
 \delta^3({\bf k}+{\bf k}'-{\bf G}_{\bm n})
 {\cal F}(k,k';{\bf G}_{\bm n}) \ .
 \end{aligned}
\label{eq:mastercovariance}
\end{equation}
Thus
\begin{equation}
 {\bf k}+{\bf k}'={\bf G}_{\bm n}
 =K\sum_{A=1}^{6}n_A\bq_A,\qquad \bm n\in\mathbb Z^6,
\label{eq:fullBragg}
\end{equation}
with the twelve first-star relations corresponding to $\bm n=\pm\bm e_A$.
At this order, higher-index elements are suppressed by the
Bessel-product hierarchy in Eq.~\eqref{eq:mastercovariance}. For a
time-dependent $\beta$, the coefficients $\delta C_{\bm n}$ must remain
inside the time integral; the constant-coefficient expressions above
apply when the coefficients vary negligibly during the relevant mode
evolution.

The benchmark does not satisfy a uniform small-contrast condition.
To check its amplitudes beyond first order in $\delta C$, we use the
long-pattern limit $K\ll p$, where $p=|\bk-\bk'|/2$.
At leading slow-roll and spatial-gradient order, the local spectrum
is proportional to $\sqrt{C(\bx)}$, with $C(\bx)=c_s^{-2}(\bx)$,
without an expansion in $\delta C$. Its Fourier coefficients, with
the known phase factors removed, are
\[
 d_{\bm n}=
 \int_{[0,2\pi]^6}\frac{d^6\theta}{(2\pi)^6}\,
 e^{-i\bm n\cdot\bm\theta}
 \sqrt{1+u\,e^{\beta(\sum_A\cos\theta_A-\Xi_0)}} .
\]
The covariance coefficient at $\bk+\bk'={\bf G}_{\bm n}$ is then
$H^2e^{i\bm n\cdot\bm\phi}d_{\bm n}/(4\epsilon\mpl^2p^3)$.
For the benchmark, the numerical analysis gives $d_{\bm e_A}=0.10686087$, compared with the first-order
value $\bar c_s u e^{-\beta\Xi_0}I_1(\beta)I_0(\beta)^5/2=0.11075321$.
The leading first-star amplitude therefore changes by $3.5\%$ in
this limit. 
The allowed momentum transfers remain in the same reciprocal module:
the spatial coefficients are periodic in the six phases, and successive
mode mixing only adds module vectors. Thus
Eq.~\eqref{eq:mastercovariance} gives first-order amplitudes, while
the momentum-transfer selection rule is preserved beyond that
approximation.

\section{Conclusions}

We have formulated quasicrystalline inflation as a rank-six icosahedral phase medium with three phonons and three phasons. We have explained relevant concepts such as the  phason.
We have taken an effective theory approach to quasicrystalline inflation. We discussed the background solution and its sustainability.
We have left the UV completion issue for future work.

We derived the quadratic action for curvature perturbations. 
We considered a pure-phason displacement as a benchmark.
We have demonstrated that a constant phason configuration gives rise to  a nonzero dipole with a vanishing linear geometrical quadrupole.
We have checked that the nonlinear contribution to the quadrupole anisotropy can be controlled to match observations.

The main falsifiable prediction in this paper is a hierarchy of off-diagonal correlations at 
\begin{equation}
 {\bf k}+{\bf k}'=K\sum_{A=1}^{6}n_A\bq_A,\qquad \bm n\in\mathbb Z^6,
\end{equation}
with leading-order amplitudes determined by the common Bessel hierarchy. The reciprocal-module selection rule is preserved beyond first order.
This is the cosmological primordial diffraction pattern.
Thus, in principle, we can provide evidence for primordial quasiperiodic order in the early universe observationally.

In this paper, we have considered icosahedral quasicrystalline inflation. It is intriguing to consider other types of quasicrystalline inflation.  It is also interesting to consider tensor perturbations in quasicrystalline inflation
and the quantum entanglement structure of primordial gravitational waves~\cite{Jiro}.

\vspace{0.1cm}
\section*{Acknowledgments}
J.\ S. was in part supported by JSPS KAKENHI Grants No. JP23K22491, No. JP24K21548, and No. JP25H02186.

\appendix

\section{Pure-phason benchmark and dipole norm}

Using the phason basis in Sec.~II B and
$w_a=w_*=\pi/\tau^2$, one has
\begin{equation}
 \bm\phi=w_*
 \left[\bm P^{(1)}+\bm P^{(2)}+\bm P^{(3)}\right],
 \qquad
 \sum_{A=1}^{6}\bm q_A\,\phi_A=0 .
\end{equation}
The phases are
\begin{equation}
 \bm\phi=(\pi\tau,\pi\tau,-\pi\tau,w_*,w_*,w_*)
 \equiv(-w_*,-w_*,+w_*,+w_*,+w_*,+w_*)\pmod{2\pi},
\end{equation}
where we used $\tau+w_*/\pi=2$ in the equivalent phase representation.  Since all cosines are equal, $C_{ij}=2\cos w_*\delta_{ij}$ and the traceless quadrupole vanishes.

For $s=(-1,-1,+1,+1,+1,+1)$,
\begin{equation}
 {\bf S}=\sum_As_A\bq_A
 =\frac1{\cal N}(2\tau+2,\,-2,\,0),
\end{equation}
and direct use of $\tau^2=\tau+1$ gives
\begin{equation}
 |{\bf S}|^2=6+\frac{6\sqrt5}{5}.
\end{equation}
Combining with Eq.~\eqref{eq:dipolegeneral} yields Eq.~\eqref{eq:centralA}.

\section{Full-module Fourier coefficients}

The generating identity
\begin{equation}
 e^{\beta\cos\Theta}=\sum_{n=-\infty}^{\infty}I_n(\beta)e^{in\Theta}
\end{equation}
applied independently to all six phases gives Eq.~\eqref{eq:Besselexpansion}.  In particular, the first-star coefficient is $I_1I_0^5$, a two-generator coefficient such as $\bm e_A+\bm e_B$ has weight $I_1^2I_0^4$, and the ratio is $I_1/I_0\simeq\beta/2$ for $\beta\ll1$.  This supplies a controlled expansion parameter for truncating the observable module while retaining a genuinely quasiperiodic parent state.

\section{Normalization of the in-in response kernel}
\label{app:inin}

Let $A(\eta)\equiv\epsilon\mpl^2a^2$ and $C({\bf x})=\bar c_s^{-2}+\delta C({\bf x})$.  The kinetic Lagrangian density is
\begin{equation}
 {\cal L}_{\rm kin}=A C\,\zeta'^2,
 \qquad
 \Pi_\zeta\equiv\frac{\partial{\cal L}}{\partial\zeta'}=2AC\,\zeta'.
\end{equation}
The Hamiltonian density is
\begin{align}
 {\cal H}
 =\,\Pi_\zeta\zeta'-{\cal L}
 =\frac{\Pi_\zeta^2}{4AC}+A(\nabla\zeta)^2 
 =\,\frac{\Pi_\zeta^2}{4A\bar c_s^{-2}}+A(\nabla\zeta)^2
 -\frac{\Pi_\zeta^2}{4A(\bar c_s^{-2})^2}\,\delta C
 +O(\delta C^2).
\end{align}
In the interaction picture, $\Pi_{\zeta,I}=2A\bar c_s^{-2}\zeta'_I$, and therefore
\begin{equation}
 {\cal H}_{\rm int}=-A\,\delta C({\bf x})\,\zeta_I'^2
 =-{\cal L}_{\rm int}+O(\delta C^2).
\label{eq:HintLegendre}
\end{equation}
Using Eq.~\eqref{eq:Fourierconv},
\begin{equation}
 H_{\rm int}=-\epsilon\mpl^2a^2
 \sum_{\bm n\neq0}\delta C_{\bm n}
 \int\frac{d^3p}{(2\pi)^3}\,
 \zeta'_{\bf p}\zeta'_{-{\bf G}_{\bm n}-{\bf p}} .
\end{equation}
With
\begin{equation}
 \zeta_{\bf k}(\eta)=u_k(\eta)a_{\bf k}
 +u_k^*(\eta)a^\dagger_{-{\bf k}},
 \qquad
 [a_{\bf k},a^\dagger_{\bf p}]=(2\pi)^3\delta^3({\bf k}-{\bf p}),
\end{equation}
the two Wick contractions give
\begin{align}
 \delta\langle\zeta_{\bf k}\zeta_{\bf k'}\rangle
 =(2\pi)^3\sum_{\bm n\neq0}
 \delta^3({\bf k}+{\bf k'}-{\bf G}_{\bm n})
 \delta C_{\bm n}
 \left\{-4\epsilon\mpl^2\,{\rm Im}
 \left[u_k(0)u_{k'}(0)
 \int_{-\infty(1-i0^+)}^0d\eta\,a^2
 u_k^{\prime *}u_{k'}^{\prime *}\right]\right\},
\end{align}
which proves Eqs.~\eqref{eq:ininmodule} and \eqref{eq:Fkernel}.

For exact de Sitter, $a=-1/(H\eta)$ and
\begin{equation}
 u_k'(\eta)=
 \frac{H\bar c_s^{\,2}k^2\eta}
 {2\sqrt{\epsilon \bar c_s}\,\mpl k^{3/2}}
 e^{-i\bar c_s k\eta}.
\end{equation}
Hence
\begin{equation}
 \int_{-\infty(1-i0^+)}^0d\eta\,a^2
 u_k^{\prime *}u_{k'}^{\prime *}
 =-\frac{i\bar c_s^{\,2}\sqrt{k k'}}
 {4\epsilon\mpl^2(k+k')},
\end{equation}
and substitution of the late-time amplitudes gives Eq.~\eqref{eq:FdeSitter}.

\bibliographystyle{unsrt}
\bibliography{quasicrystalline_inflation}

\end{document}